# I, Quantum Robot: Quantum Mind control on a Quantum Computer

P. A. Zizzi

Dipartimento di Matematica Pura ed Applicata
Via Trieste, 63, 35121 Padova, Italy

zizzi@math.unipd.it

## Abstract

The logic which describes quantum robots is not orthodox quantum logic, but a deductive calculus which reproduces the quantum tasks (computational processes, and actions) taking into account quantum superposition and quantum entanglement.
A way toward the realization of intelligent quantum robots is to adopt a quantum metalanguage to control quantum robots. A physical implementation of a quantum metalanguage might be the use of coherent states in brain signals.

## 1. Introduction

Quantum robots were first introduced by Benioff [1]. The original aim was to give a complete validation of quantum mechanics, by means of quantum systems, which, (like quantum robots) might carry out both theoretical calculations (quantum computing) and experiments.
A quantum robot can, in principle, perform a self-measurement: computational steps looked from the robot's point of view are quantum measurements of its own state. This was already recognized by Benioff, who also pointed out the difficulty of self-reference arising in this case. This in fact requires a new, quantum approach [2] to Gödel's incompleteness theorem.
Benioff defined a quantum robot as a mobile system which has a quantum computer on board, and any needed ancillary systems.
The quantum robot moves in and interacts with the environment of a quantum system. Environments are taken to be systems in discrete space lattices for simplicity. The quantum computer on board can be described by a quantum Turing machine (QTM) [3], quantum networks [4], and quantum cellular automata (QCA) [5].
The dynamics of a quantum robot and its interactions with the environment is described in terms of *tasks*. A task for a quantum robot is equivalent to a quantum function for a quantum computer. Each task consists of a sequence of computation and action phases. The computational phase is aimed to determine what action (or move) the quantum robot should make.
The quantum robots originally discussed by Benioff have no awareness of their environment, and do not make decisions or measurements. However, in the future it might be that quantum robots will be aware of the environment, and could perform experiments.
A quantum robot which interacts with a classical environment, will suffer decoherence. Benioff gave a solution to this problem by introducing alternating phases of (quantum) computation and (classical) functioning. This alternation of phases reminds the classical and quantum configuration phases of tubulins in the Penrose-Hameroff model of quantum mind [ 6]. This suggests the possibility of studying a quantum cyborg (a quantum mind interacting with a quantum system capable of quantum computation), whose alternating phases are synchronized with those of tubulins. However, in this paper, we will not discuss this particular model. Instead, we will investigate about a quantum cyborg defined as a quantum control coming from the brain in terms of generalized coherent states [7]. This quantum control acts as a trigger of quantum computation, by providing at least, one qubit to a quantum system able to support this kind of computation (like, for example quantum dots [8]).
These generalized coherent states are very robust against dissipation phenomena in the brain [9 ]. This mechanism allows the human mind to receive a feedback from the quantum computer: the mind does not only implement quantum computation in the external quantum system, but also can follow the quantum computational process. The quantum robot is in fact, in this case, the whole ensemble of a quantum mind and a quantum computer. Also, this quantum robot might undergo decoherence due to the interaction with the external environment. However, every time this happens, (e.g., the original qubit decoheres in classical bits) then the quantum control from the brain re-supplies the quantum dots with a new qubit.

Humans should adopt a quantum metalanguage to control quantum robots. A quantum metalanguage is a metalanguage which reflects properly into a quantum object-language, that is, in the logic of quantum computers. The reflection principle between (classical) metalanguage and object language was introduced in [10] and then extended to the quantum case in [11].
The terms "self- awareness" and "consciousness" will be used in this paper in the sense that the quantum robot under consideration is in fact a quantum cyborg, that is, a quantum computer controlled by a human (quantum) mind via quantum metalanguage.

The possibility that quantum robots might, in principle, develop, by some self-organization mechanism, their own quantum logic, which will be inaccessible to humans, looks very unlikely. In fact, the quantum cyborg is the ensemble human brain-quantum computer, and any development/modification in the quantum computer's logic should be accessible to the human mind. Eventually, the terms "self- awareness" and "consciousness" used in this paper, when referred to quantum robots, are not derived from the proposed quantum metalanguage, but are just assumed, because in this case the quantum control is a consequence of quantum processes in the human brain.

## 2. The logic of a quantum robot

In [12], unitary transformations were given the status of generalized quantum measurements [13]. In this way, quantum logic gates, which perform quantum computation, can also be reinterpreted as internal, reversible generalized quantum measurements if an internal observer is allowed. The internal observer can be interpreted in (at least) four different ways:

i) An observer in a quantum space which is in a one-to-one correspondence with the state space of the quantum system [14]. Of course, this view is possible only in the framework of (loop) quantum gravity [15].

ii) An observer in a classical space, who is, however, capable to communicate with the logic of the quantum computer by means of a quantum metalanguage, which is a suitable quantum control.

iii) The quantum robot. Of course, the quantum robot is able to look at its own computation via internal measurements (unitary transformations) without destroying quantum superposition.

Also, the quantum robot can perform quantum measurements on other quantum systems, comprising other quantum computers. From outside, the quantum robots performing measurements on other quantum systems of the environment, is viewed, by the external observer, as a generalized quantum measurement.

iv) A human being with a QCA implanted in his brain that is, a quantum computing cybernetic organism (Quantum-Computing Cyborg).

The quantum logics of cases i) iii) and iv) are reproducible in terms of networks of quantum logic gates, measurements are all internal and are described by unitary operators (or at least by generalized measurements in the case of a quantum robot performing an experiment, a particular subcase of case iii)). Instead, case ii) needs a quantum metalanguage.
The premises of the logical calculus are the assertions of the quantum metalanguage, which, hopefully, can be physically implemented by coherent state inputs from brain signals, as we will see in the next section. Before going into further detail, we wish to stress the fact that quantum metalanguage, intended as a quantum control, can be viewed as an expression of intentionality, high-level decision making, and in general high-level thought (in the following, we will relate quantum metalanguage to metathought). To us, in any case, intentionality and consciousness seem to have an important feature in common, namely, they are not directly testable by an external observer without being destroyed (like in the quantum measurement scheme). This suggests that consciousness and intentionality (as well as other high-level forms of thought ) belong to the quantum domain.
We are looking for the most adequate logic for a quantum robot. To this aim, we will remind some notions about logical propositions and assertions in orthodox quantum logic [16], and we will illustrate why they do not fit the quantum computational case.
In orthodox quantum logic a proposition is interpreted as a projector. In two-dimensional complex Hilbert space, there are two projectors, $P_0$ and $P_1$. More precisely, the two atomic propositions $p_0$ and $p_1$ are, in the Hilbert interpretation *H*:

$$p_0{}^H = P_0\,, \quad p_1{}^H = P_1 \qquad (2.1)$$

The eigenvectors of $P_0$ and $P_1$ are $|0\rangle$ and $|1\rangle$ respectively, with real eigenvalue 1:
$P_0|0\rangle = |0\rangle$, $P_1|1\rangle = |1\rangle$.
Then, the states $|0\rangle$ and $|1\rangle$ can be taken as the representative of the propositions $p_0$ and $p_1$ as such propositions are asserted (true) at those states.
The deductive logical calculus of quantum computers cannot be developed in the framework of orthodox quantum logic. In fact, a quantum computer is not just a quantum system, but a quantum system which computes, and the logical calculus should reproduce the quantum computational process. A quantum robot is even more complex: it is a quantum system, which can perform measurements, and it is also a quantum computer which computes its own moves. The whole system can, in turn, be simulated by another quantum computer. In the internal logic of a quantum computer (or of a quantum robot) the syntax and the semantics are strictly inter-twined, which leads to a kind of self-reference. However, the latter is a problem only within a classical metalanguage; in the case of a quantum metalanguage, self-reference can be formalized as a metatheorem [17].
Orthodox quantum logic can describe only projective measurements, which have the effect of destroying quantum superposition, stopping the quantum computational process, and losing information.
Orthodox quantum logic is then the antithesis of quantum information and quantum computation (sequent calculus should reproduce the computational process, instead of stopping it). Moreover, orthodox quantum logic is structural: there are the two structural rules of contraction and weakening which disagree [18] with the two no-go theorems of quantum computing, namely the no-cloning [19] and the no-erase [20] theorems, respectively.
We will briefly give the definitions of sequent and assertions, and metalanguage, which are very important for what follows.
The following notation will be used:
The symbol $|-$ known as the turnstile separates the *assumptions* on the left from the *propositions* on the right.
*A* and *B* denote formulae of first-order predicate logic (one may also restrict this to propositional logic). Γ,Δ,Σ, and Π are finite (possibly empty) sequences of formulae, called contexts.
When on the *left* of the $\vdash$, the sequence of formulas is considered *conjunctively* (all assumed to hold at the same time), while on the *right* of the $\vdash$, the sequence of formulas is considered *disjunctively* (at least one of the formulas must hold for any assignment of variables).
We say that the finite list Δ of assertions Bj (j =1,2,….k) *follows from* a finite list Γ of assertions Ai (i=1,2….n) (or equivalently Γ *yields* Δ) and write: $\Gamma |- \Delta$. Where $|-$ ("yields" or "therefore") is a metalinguistic link between assertions. $\Gamma$ is said the antecedent, and Δ the consequent of the sequent.
The symbol $|-$ is said "turnstile".
The sequent $\Gamma |- \Delta$ can be rewritten, more extensively, as:
$A_1,...,A_n |- B_1,....B_k$
Where $A_i$ and $B_j$ $(i = 1,2...,n)$ $(j = 1,2...,k)$ are assertions.
Either $\Gamma$ or $\Delta$ (or both) can be empty.
If the consequent $\Delta$ is empty: $\Delta \equiv \emptyset$ this is interpreted as false, that is, $\Gamma |-$ means that $\Gamma$ proves falsehood, and therefore it is inconsistent.
Instead, an empty antecedent $\Gamma \equiv \emptyset$ is assumed true, that is, $|- \Delta$ means that $\Delta$ follows without any assumption, that is, it is always true.
We say then that $|- \Delta$ is a logical assertion
In the case of projectors as propositions, the assertions are denoted, in terms of sequents, by:

$|\!-p_0$ and $|\!-p_1$. (2.2)

In a abstract generalization, "M-algebras" [21], which does not forcedly consider Hilbert spaces, propositions are operators acting on physical states. The most important requirements, for an operator to be viewed as a proposition, is that it must be hermitian and idempotent (which, in the Hilbert case corresponds to projectors). We interpret the above restrictions as follows. Hermitian operators have real eigenvalues. In particular, idempotent operators have eigenvalues 0 or 1, that is, they allow for asserting or negating in the classical way. When the operator is not hermitian, it is true that there is no way to interpret it directly as a logical proposition, because its eigenvalues are not real numbers, and the proposition cannot be asserted as usual. The problem is the lack of generalization: when the operators are not hermitian and idempotent, there is really no way to formulate propositions and make assertions on a physical system? There is not a no-go theorem which prevents defining a complex assertion degree, and a truth degree as the squared modulus of the assertion degree.

We overcome this problem by taking non hermitian operators acting on the basis states of the Hilbert space, as atomic propositions, and the complex eigenvalues as degree of assertion of such propositions.

In quantum computational logics [22] propositions are interpreted as qubit states. In that case, the semantics of a proposition is not the truth, but information. Instead, in the framework of non hermitian operators discussed here, the semantics is given in terms of quantum coherence and quantum probability..

We give now a short review of the reflection principle, introduced in the framework of basic logic [10], by which the metalanguage reflects into the object language.

**Reflection Principle**: All the connectives of Basic logic satisfy the principle of reflection, that is, they are introduced by solving an equation (called *definitional equation*), which "reflects" meta-linguistic links between assertions into connectives between propositions in the object-language:

| | | | |
|---|---|---|---|
| **Obiect-language** | $\xleftrightarrow[\text{Re}\,flection]{}$ | **Metalanguage** | |
| Logical connectives | $\xleftrightarrow[\text{Re}\,flection]{}$ | Metalinguistic links $|\!-$; <u>and</u> | (2.3) |
| (Btween propositions) | | (Between assertions) | |

**Example**. Definitional equation for the (classical) connective & = "and"):

$\Gamma|\!-A\,\&\,B$ <u>iff</u> $\Gamma\ |\!-\ A$ <u>and</u> $\Gamma|\!-B$ (2.4)

Where the antecedent $\Gamma$ can be empty.

### 3. A possible physical implementation of quantum metalanguage

Case ii, mentioned in the previous section, which corresponds to a non-invasive approach to BCI, could be physically implemented in an analogous way to the classical case, by sending electromagnetic waves from the brain to the robot. In the quantum case, however, electromagnetic waves should be in a coherent state generated by a quantum process. But the brain does not always produce such coherent states. The problem arises by the fact that signals emitted by the brain are a mixture of different electromagnetic waves. In fact, these signals are coming from a number of different sources, some of which operating in a noisy fashion, while others undergo quantum processes. Therefore, the selection of coherent states within this mixture would require a very sophisticated technology, and a very high level of accuracy.

We remind that a coherent state is a particular kind of quantum state of the quantum harmonic oscillator.

The coherent state $|\alpha\rangle$ is a right eigenstate of the annihilation operator *a*:

$$a|\alpha\rangle = \alpha|\alpha\rangle \quad (3.1)$$

It should be noticed that the annihilation operator *a* is not hermitian, then its eigenvalue $\alpha$ is a complex number:

$$\alpha = |\alpha| e^{i\theta}, \text{ with : } \theta \in R \tag{3.2}$$

Where $|\alpha|$ and $\theta$ are called the amplitude and the phase of the coherent state, respectively.

Physically, this means that a coherent state is left unchanged by the annihilation of a particle.
It should be noticed that coherent states satisfy a closure relation, which is the resolution of identity:

$$\frac{1}{\pi}\int |\alpha\rangle\langle\alpha| d^2\alpha = 1 \tag{3.3}$$

Then, any state can be decomposed on the set of coherent states. That is, coherent states can be selected out from other states. This would be very important, of course, in the case of brain signals, which should be the inputs for a QCA.
In what follows, we show why coherent states are of importance for quantum computation, in particular for quantum control. The latter can be put in a axiomatic form to provide a formal quantum metalanguage (the adequate metalanguage for the logic of a quantum computer).
In the case of the annihilation operator, we take as atomic propositions its right eigenstates, the coherent states $|\alpha\rangle$, and as degrees of assertions, the complex eigenvalues $\alpha$.

In this case, an atomic proposition $p$ is, in the Hilbert interpretation:

$$p_\alpha{}^H = |\alpha\rangle. \tag{3.4}$$

An assertion is denoted by: $\vdash^{\alpha} p_\alpha$, where the superscript $\alpha$ on the sequent symbol is the assertion degree which, in the interpretation, corresponds to a probability amplitude.
In [11] the assertion degree *g* of an atomic proposition *p* interpreted as an eigenstate of a non-hermitian operator O, was defined as the complex eigenvalue:

$$g \equiv \langle p^H | O | p^H \rangle. \tag{3.5}$$

Moreover, the truth degree *v* was defined by:

$$v \equiv \langle p^H | O^\dagger O | p^H \rangle = |g|^2. \tag{3.6}$$

In the case under consideration the assertion degree of the proposition $p_\alpha$ is:

$$g \equiv \langle \alpha | a | \alpha \rangle = \alpha. \tag{3.7}$$

And the truth degree is:

$$v \equiv \langle \alpha | a^\dagger a | \alpha \rangle = |\alpha|^2. \tag{3.8}$$

It should be noticed that the proposition $|\alpha\rangle$ is true, but not in the classical sense, that is, its truth values are not just $\{0,1\}$, but are in the range $[0,1]$ like in many-valued logic and fuzzy logic [23].

The proposition has a truth-degree: $|\alpha|^2 \in [0,1]$.

The truth degree of the proposition $|\alpha\rangle$ is equal to the average photon number in the coherent state:

$$\langle n \rangle = \langle a^\dagger a \rangle = |\alpha|^2. \tag{3.9}$$

Two different coherent states, $|\alpha\rangle$ and $|\beta\rangle$ are not orthogonal:

$$\langle \beta | \alpha \rangle = e^{-\frac{1}{2}\left(|\beta|^2 + |\alpha|^2 - 2\beta^*\alpha\right)} \neq \delta(\alpha - \beta) \tag{3.10}$$

Therefore, if the quantum oscillator is in the coherent state $|\alpha\rangle$, it is also with nonzero probability in the other coherent state $|\beta\rangle$.

This means that two assertions $\vdash^{\alpha} p_\alpha$, and $\vdash^{\beta} p_\beta$ can be done simultaneously, or in more precise logical terms, we can make the juxtaposition:

$$\vdash^{\alpha} p_{\alpha} \quad \underline{\text{and}} \quad \vdash^{\beta} p_{\beta} \tag{3.11}$$

Where <u>and</u> is the metalinguistic link between the two assertions.
In our case, the metalanguage in (3.11) should be quantum, as it concerns two inputs from the brain signals in terms of coherent states.
This quantum metalanguage will reflect in the object-language of the QCA, in terms of a quantum superposition:

$$\vdash p_{\alpha}({}_{\alpha}\&_{\beta})p_{\beta} \quad \underline{\text{iff}} \quad \vdash^{\alpha} p_{\alpha} \quad \underline{\text{and}} \quad \vdash^{\beta} p_{\beta} \tag{3.12}$$

Where ${}_{\alpha}\&_{\beta}$ = "connective of quantum superposition", is defined as:

$$p_{\alpha}({}_{\alpha}\&_{\beta})p_{\beta} \equiv (\alpha p_{\alpha})\,\&\,(\beta p_{\beta}) \tag{3.13}$$

Where & stands for the (classical) logical connective "and".
The quantum connective ${}_{\alpha}\&_{\beta}$ represents the quantum superposition of two (coherent) states:

$$\alpha|\alpha\rangle + \beta|\beta\rangle \tag{3.14}$$

The quantum metalanguage in (3.11) was introduced by the use of coherent states. However, once one has obtained the quantum logical connective ${}_{\alpha}\&_{\beta}$ for the object-language in terms of a superposition of coherent states, it is possible to select from the latter the qubit state as follows.
The representation of the coherent state $|\alpha\rangle$ in the basis of Fock states is:

$$|\alpha\rangle = e^{-\frac{|\alpha|}{2}\sum_{n=0}^{1\infty}\frac{\alpha^{n}}{\sqrt{n!}}|n\rangle} \tag{3.15}$$

For $n = 0,1$ we have, from (3.15):

$$|\alpha\rangle = e^{-\frac{|\alpha|^{2}}{2}}\left(|0\rangle + \alpha|1\rangle\right) \tag{3.16}$$

Then, the quantum superposition of two coherent states in (3.14) can be rewritten as the qubit state

$$\lambda_{0}|0\rangle + \lambda_{1}|0\rangle \tag{3.17}$$

With:

$$\lambda_{0} = e^{-\frac{|\alpha|^{2}}{2}} + e^{-\frac{|\beta|^{2}}{2}}, \qquad \lambda_{1} = \alpha e^{-\frac{|\alpha|^{2}}{2}} + \beta e^{-\frac{|\beta|^{2}}{2}}. \tag{3.18}$$

With the constraint: $|\lambda_{0}|^{2} + |\lambda_{1}|^{2} = 1$. (3.19)

The above constraint, of course imposes constraints on $\alpha$ and $\beta$ in the metalanguage. A constraint on a metalanguage, which provides the inputs or data, is referred to as "metadata" (data about data).

**4. Quantum metalanguage and self-awareness**

If consciousness is really a product of some quantum effects occurring in our brain], quantum computers controlled by a quantum mind might also be considered "awake", and we should figure out what might happen in this case.
To this aim, we remind some important features of quantum information:
i) Quantum superposition: The qubit can be in the superposed state $\alpha|0\rangle + \beta|1\rangle$.
ii) Quantum entanglement: A bipartite quantum state cannot be written as a tensor product. In a sense, the two original qubits lose their individuality.
iii) A two-qubit maximally entangled state (Bell state) is: $\frac{1}{\sqrt{2}}\left(|0\rangle_{A}\otimes|0\rangle_{B} + |1\rangle_{A}\otimes|1\rangle_{B}\right)$.
iv) No cloning: It is not possible to copy an unknown quantum state.
v) No erase: It is impossible to delete an unknown quantum state.
vi) We cannot measure a qubit in its wholeness; in fact in a quantum measurement we always lose information.
It is not difficult to imagine what happens when all that applies to a quantum robot.

“He” can get entangled with other quantum robots.
We cannot erase its own information, whatever it is.
If we want to get some information, we should destroy the robot.
The most important features of quantum computation are:
Massive quantum parallelism (due to quantum superposition and entanglement) leading to computational speed-up. (Quantum computers can perform some computational tasks exponentially faster than classical computers).
Reversibility: A computational step can be undone. This is due to the fact that the quantum logic gates are mathematically described by unitary operators.
One can figure out then, what happens when a quantum robot that is “awake” processes its own “thought”.
Let us see what the effect of massive parallelism on proofs is.
We give the quantum robot a theorem *T* very difficult to prove. He is very efficient and does the task very fast.
He gives us the result, for example “true”, but not the proof.
He would say: “You cannot understand that”, and he will be right, in facthe proof is “superposed” , therefore we will know only the result of his “thought” process, namely a classical bit.
Moreover, a quantum robot that is “awake”, will use a kind of reversible thought. For example “he” can think something**,** and then pretend he never did. But he is not a “liar”; he just completely forgot his previous thought. In this case, there will be no information available to us about his past thoughts.
Quantum robots can get entangled with other robots, and act together at a distance.
On the other hand, a quantum robot would never be able to clone itself, and could not be cloned either: he will be an individual, differently from classical robots.
Interaction between intelligent agents communicating via entangled resources has been explored under the domain of quantum game theory [24 ] [25 ] [26 ].
We believe that the quantum robots discussed in this paper, encompassing both intelligent agents and quantum computers, might play quantum games as well.

## 5. Quantum metalanguage-control and domains of metathought

We will refer to the metalanguage as a control over the object language [27], in computer science, with particular attention to quantum computing. In the scheme of control an important role is played by representation of the controlled system in the controlling device.
The language of the controlled system is referred to as the object-language *L*. The language used to make the representations in the controlling system, let it be *ML*, is referred to as a metalanguage.
A feedback control is checking up on the state of the system. In every feedback loop, information about the result of a transformation or an action is sent back to the input of the system in the form of input data. These new data facilitate and accelerate the transformation in the same direction as the preceding results, they are positive feedback - their effects are cumulative.
If the new data produce a result in the opposite direction to previous results, they are negative feedback - their effects stabilize the system. In the first case there is exponential growth or decline; in the second there is maintenance of the equilibrium.
In terms of metalanguage and object-language, there is a feedback from the object-language to the metalanguage. If the feedback is a positive loop, the logical theory “explodes”.
It would correspond to an object-language containing its own semantics.
We need therefore a control in terms of negative loops.
This is very difficult when the object-language is quantum, as there is no way to control the positive feedback loops by a classical control (a classical metalanguage). We need therefore a quantum control (given in terms of a quantum metalanguage).
To date, there is no way to get a feedback control of a quantum system without performing a measurement, which changes the quantum state: this is referred to as back action. The effect of back action is to introduce quantum noise. In the worst case (projective measurements) the feedback control causes the instantaneous collapse of the quantum state. In the less invasive case (weak measurements [28]) there is a less abrupt reduction of the quantum state. When the quantum system is a quantum computer, the computational process is stopped in the first case, and is strongly disturbed in the second one. We believe that a quantum metalanguage of the kind described in the previous sections will be the most adequate quantum control for a quantum robot. In fact, in this case the feedback control can check up on the state of the system practically from inside, without any back action.
The QCA might self-organize in such a way to display a stronger control on the quantum metalanguage-control itself. The power stands in the control. Let us make an example. Let us consider a QCA which self-organizes to reach an apparent "self-awareness" but which is still incapable of performing a quantum control.
An apparently "self-aware" quantum robot would say: "I am", knowing exactly what that means. That would appear as the best realization of strong AI to some people who use a classical metalanguage, but not to others who use a quantum metalanguage.
In fact, knowing the meaning of self-awareness, even in its substance, it is not the same of being really self-aware. The difference stands in metathought, the mental process of thinking about our own thought**.** The term metathought was introduced by Perry [29], who proposed that individuals with more sophisticated epistemological beliefs were more likely to engage in personal reflection and analysis about their understandings and use of knowledge. Metathought is the process of thinking about thinking.
However in the literature this concept is better known as metacognition, a term introduced by Flavell [30]. Flavell describes metacognition as follows: "Metacognition refers to one's knowledge concerning one's own cognitive processes or anything related to them…". Metacognition represents the "executive control" system in many cognitive theories. Metacognition, or the ability to control one's cognitive processes (self-regulation) has been linked to intelligence by Sternberg, [31]. Sternberg refers to these executive processes as "metacomponents" in his theory of intelligence. Metacomponents are executive processes that control other cognitive components as well as receive feedback from these components. According to Sternberg, metacomponents are responsible for "figuring out how to do a particular task or set of tasks, and then making sure that the task or set of tasks are done correctly".
Here we use the older term "metathought" in a more general, theoretical sense, which seems to be more adequate in this context, as it does not deal forcedly with practical applications in learning, memory, problem solving etc.
Metathought can be applied to logic: thinking about our thinking about a metalanguage, which in turn speaks about an object-language.
Therefore, metathought allows choosing the most appropriate metalanguage which reflects properly in the object-language under consideration. This is illustrated in the following commutative diagram:

$$\begin{array}{ccc} Thought & \xrightarrow{f} & Object\ language \\ g\uparrow & & \uparrow h \\ Metathought & \xrightarrow{k} & Metalanguage \end{array} \qquad h \circ f = k \circ g \qquad (5.1)$$

Where *h*, *f*, *g*, *k*, are functors which relate different categories.
If the object-language is quantum, the metalanguage must be quantum too. Let us consider for example a classical metalanguage which reflects (improperly) in a quantum object-language:

$$\vdash p_0 ({}_{\alpha}\&_{\beta}) p_1 \quad \underline{\text{iff}} \quad \vdash p_0 \quad \underline{\text{and}} \quad \vdash p_1 \qquad (5.2)$$

The definitional equation in (5.2) is evidently wrong: Where the probability amplitudes $\alpha$ and $\beta$ ( in the object-language on the LHS) come from, if they don't appear in the metalanguage on the RHS? On the other hand, the reflection of a quantum metalanguage in a classical object-language is wrong as well.

$$\vdash p_0 \,\&\, p_1 \quad \underline{\text{iff}} \quad \vdash^{\alpha} p_{\alpha} \quad \underline{\text{and}} \quad \vdash^{\beta} p_{\beta} \tag{5.3}$$

In fact, the assertion degrees $\alpha$ and $\beta$ in the metalanguage on the RHS of the definitional equation in (5.3) disappear in the object-language on the LHS.
Therefore, metathought is a property of a controlling agent, who has to decide what the most adequate metalanguage is. With opportune boundary conditions, an apparently self-aware quantum robot reaches the level of thought. In this case the robot can be still controlled by a quantum metalanguage which prevents him to reach the level of metathought. Metathought can be formalized as a metatheory of thought.
The theories of meta-knowledge which are related to different common properties of a selected class of theories are metatheories.
A meta-theory M may represent the specific point of view on a certain class or set of theories T and this viewpoint generates meta-properties of T**.** Meta-properties are the consequence of the relation between M and T**,** but they are not the properties of any T application domain.
A metatheory [28 ] is the theory of a domain DT**,** where:

$$DT = T_1 \cap ...T_{k-1} \cap T_k \tag{5.4}$$

With a common domains D**,** and it may refers to the pre-selected class of properties of DT (selection of the viewpoint).
If metathought is metatheory of thought, however, it might be possible that a perturbation due to back action of the QCA might lead to the collapse of metathought to the common domain of thought. Let us consider, as an example, magnetic domains of ferromagnetic materials. The domains are randomly aligned; however, the presence of a magnet makes the domains gradually align with each other. In this way, the ferromagnetic material becomes a permanent magnet itself. Magnetic domains randomly aligned may represent metathought, all domains aligned in the same direction represent thought, and the magnet M represents the disturbance (the back action of the QCA on the quantum metalanguage). The paradigm of collapse of metathought to thought is illustrated below:

↑↑↑ ↓↓ ↑↑ ↓↓↓ ↑↑↑↑ Metathought

⇓ back-action of QCA (5.5)

↑↑↑↑↑↑↑↑↑↑ ↑↑↑↑ Thought

**6. Concluding remarks**

A quantum computer is not simply a quantum system, but a quantum system which computes: quantum computation is a quantum process. Then, the ontology of quantum computation is mainly ontology of processes [32] as in the case of Quantum Field Theory (QFT). This puts quantum computation on the border between Quantum Mechanics and QFT. The ontology of processes of QC, which is given in terms of probability amplitudes, reflects, in logic, in a quantum metalanguage, and, in turn, in an object-language which is different from orthodox quantum logic. Another suggestion of a possible relation between QC and QFT comes from the discussion of coherent states in brain signals for control of a QCA. In fact, those considerations showed up the necessity of second quantization in terms of annihilation and creation operators. Quantum computing on its own is related to QFT only by the same kind of ontology. Instead, a quantum robot is more closely related to QFT because its control comes from the quantum electromagnetic field. In a sense, a physical implementation of quantum metalanguage is the starting point toward a QFT- computing.

**Acknowledgments**
I am very grateful to E. Pessa  for many  useful and enlightening discussions.

.